\documentclass{icrc2009}

\usepackage{graphicx}   
\usepackage[caption=false]{caption}    
\usepackage[font=footnotesize]{subfig} 
\usepackage{fixltx2e}
\usepackage{url}

\newcommand{\shorttitle}[1]%
{\markboth{Proceedings of the 31\MakeLowercase{$^{st}$} ICRC, {\L}\'{o}d\'{z} 2009}{#1} }
\newcommand{\etal}{\MakeLowercase{\textit{et al. }}} 

\begin{document}
\title{VERITAS Observations of X-ray Binaries}

\author{\IEEEauthorblockN{Roxanne Guenette\IEEEauthorrefmark{1} for the VERITAS collaboration\IEEEauthorrefmark{2}}
\\
\IEEEauthorblockA{\IEEEauthorrefmark{1}Department of Physics, McGill University, H3A 2T8 Montreal, Qc, Canada (guenette@physics.mcgill.ca)}
\IEEEauthorblockA{\IEEEauthorrefmark{2}see R.A. Ong et al. (these proceedings) or http://veritas.sao.arizona.edu/conferences/authors?icrc2009}}

\shorttitle{R.Guenette \etal VERITAS Observations of X-ray Binaries}
\maketitle

\begin{abstract}



X-ray binaries stand as the brightest X-ray sources in the galaxy, showing both variable X-ray emission and extreme flares. Some of these systems have been recently discovered to be TeV gamma-ray emitters, with the high energy emission posited as resulting from particle acceleration in relativistic jets or from shocks between pulsar and stellar winds. VERITAS, an array of four 12m imaging atmospheric Cherenkov telescopes has accrued more than 100 hours of observation time on X-ray binaries. Here we present the results of observations on 3A 1954+319, XTE J2012+381, 1A 0620-00, EXO 2030+375, KS 1947+300, SS 433, Cygnus X-1 and Cygnus X-3.
  \end{abstract}

\begin{IEEEkeywords}
Gamma rays, observations, X-ray binaries 
\end{IEEEkeywords}
 
\section{Introduction}

X-ray binaries are X-ray emitting systems composed of a compact object, either a neutron star or a black hole, and a stellar companion. The accretion of matter from the companion star onto the compact object is mostly responsible for the X-ray emission, although some are powered by the fast rotation of the neutron star. These systems can be subdivided in two groups: High Mass X-ray Binaries (HMXB) and Low Mass X-ray Binaries (LMXB). The first group includes systems in which the companion is a massive star and where the mass transfer is due to the decretion disk of the donor star, by strong stellar wind or Roche-lobe overflow. They are bright persistent X-ray emitters and show significant variability. The second group is composed of systems containing a companion star of low mass and where the mass transfer occurs via Roche-lobe overflow. Most of them are X-ray transients. An interesting feature of some X-ray binaries (HMXBs and LMXBs) is the presence of relativistic jets; these systems are classified as microquasars.\\

The detection of TeV emission from PSR B1259-63 \cite{Aharonian2005PSRB1259}, LS 5039 \cite{Aharonian2005LS509} and LS I +61 303 \cite{Albert2006LS1,Acciari2008} has validated that efficient particle acceleration occurs in binary systems. These three objects have high-mass O or B star companions, eccentric orbits and they have all been detected in the radio (which is not common for HMXBs). The emission seems to be modulated with the orbital periods of the binary systems. In PSR B1259-63, the maximum gamma-ray flux is observed near periastron; for LS 5039, it is at inferior conjunction and for LS I +61 303, it is near apastron.\\

Two different models have been proposed to explain the TeV emission in these objects. In the first model, the emission is powered by the interaction between the pulsar wind and the stellar wind. Particles are accelerated at the shock between the winds \cite{Dubus2008,Sierpowska2008}. In the second model, which applies to the microquasar scenario, particles are accelerated in the jets \cite{Bottcher,bbb,Romero2005,Bosch}. For both models, the accelerated particles can be protons or electrons. The very-high-energy (VHE) gamma-rays are produced (in the hadronic model) by the decay of neutral pions, or (in the leptonic model) by inverse Compton up-scattering of stellar UV photons from the companion star by the relativistic electrons. The model of the shock-powered emission from the pulsar wind interaction is consistent with PSR B1259-63 observations, whereas the cases of LS 5039 or LS I +61 303 are still unclear, mainly due to the fact that the nature of the compact objects is unknown. Therefore, both models described above are plausible to explain the TeV emission. The MAGIC collaboration has reported evidence of VHE gamma-ray emission from the microquasar Cygnus X-1 \cite{MAGIC_CygX1}. If the detection is confirmed, the validity of the microquasar model would be unambiguously demonstrated.\\

X-Ray binaries are intriguing systems. From the three confirmed detections of VHE gamma-ray emission (PSR B1259-63, LS 5039 and LS I +61 303), it is clear that each system is unique. The fact that the maximum emission flux does not occur at the same orbital phase and that a correlation with the X-ray emission is not present in all three systems makes it difficult to predict when to observe these objects in the TeV range. Moreover, mechanisms involved in the VHE gamma-ray production could be initiated or enhanced during intense X-ray flares. Magnetic fields could also play a role in the production of VHE emission. Some authors predict detectable TeV emission from binaries formed by a Be star and a highly magnetised neutron star during a major X-ray outburst \cite{Orellana2007}. Finally, the gamma-ray binary scenario might resolve the unidentified nature of some TeV sources, in particular HESS J0632+057 \cite{HESS,Gernot}.\\

  \begin{table*}[th]
  \caption{Characteristics of the X-ray binaries presented (see section \textit{Results and Discussion} for references).}
  \label{table1}
  \centering
  \begin{tabular}{|c|c|c|c|c|c|c|c|}
  \hline
   Object         &  System type\footnotemark[1] & Distance (kpc) &  P$_{orb}$ (d)  & M$_{compact}$ (M$_{\odot}$) & presence of jets   \\
   \hline                                                                                                         
   Low Mass X-ray Binaries &                 &                 &                 &                              &         \\
   \hline                                                                                                         
    3A 1954+319   & M5 III + NS (P = 5.09h)&     1.7        &   $>$ 400?      &           1.4                 &  no               \\
    XTE J2012+381 & faint red star + BH?   & 3 $<$ d $<$ 12 &   ?             &            10?                &  no?              \\
    1A 0620-00    & K5 V + BH              & 1.05$\pm$0.4   &  0.323 (7.75h)  &            11                 &  yes?             \\
  \hline                                                                                                          
   High Mass X-ray Binaries &                 &                 &                 &                              &       \\
  \hline                                                                                                          
    EXO 2030+375  & B0 Ve + NS (P = 42s)   &     7.1       &   46.02          &            1.4                &   no            \\
    KS 1947+300   & B0 Ve + NS (P = 17.7s) & $\approx$10   &   40.415         &            1.4                &   no            \\
    SS 433        & A super-giant + BH?    &      5.5      &    13.1          &            9                  &   yes           \\
    Cygnus X-1    & O9.7Iab + BH           &  2.2$\pm$0.2  &    5.6           &        20$\pm$8               &   yes           \\
    Cygnus X-3    & WNe + BH?              &      9        &    0.2           &        $\leq$ 3.6             &   yes           \\
  \hline
  \end{tabular}
  \end{table*}

Eight X-ray binary systems are presented here: 3A 1954+319, XTE J2012+381, 1A 0620-00, EXO 2030+375, KS 1947+300, SS 433, Cygnus X-1 and Cygnus X-3. The first three are LMXBs; the remainder are HMXBs. SS 433, Cygnus X-1 and Cygnus X-3 are microquasars. The main characteristics of these objects are given in Table \ref{table1}. More details on these sources are given in the section \textit{Results and Discussion}.\\

\footnotetext[1]{NS: Neutron Star, BH: Black Hole P: Neutron Star Period}

\section{Observations and Analysis}
 
VERITAS is an array of four 12m imaging atmospheric Cherenkov telescopes located in Southern Arizona, USA. It is designed to observe VHE gamma-ray emission between 100 GeV and 30 TeV. The high sensitivity of the experiment allows the detection of sources with a flux of 1$\%$ of the Crab Nebula in less than 50 hours of observations. For more details see \cite{Acciari2008}.\\

During the last two years of full operation, VERITAS has accumulated more than 1600 hours of observations on many different sources. Regarding X-ray binaries, VERITAS has observed LS I +61 303 for almost 70 hours, including several hours of multiwavelength campaigns. Recent results on LS I +61 303 are presented at this conference \cite{Jamie_LSI}. Most of the data presented here were acquired in observations dedicated to other targets; the binaries were not the primary targets. The analysis has been performed using data where X-ray binaries of interest were less than 1.5$^{\circ}$ away from the pointing position.\\

For this analysis, only runs passing quality cuts (good weather and no hardware problems) were selected. The data were taken with three or four telescopes. The analysis was performed using the standard second-moment parametrisation of the telescope images \cite{Hillas1985}. Requirements are: a minimum of 500 digital counts ($\approx$ 94 photoelectrons) in the image, less than 10$\%$ of the image in the pixels at the edge of the camera and the image centre of gravity less than 1.5$^{\circ}$ from the camera centre. The following cuts were applied: -1.2 $<$ reduced mean scale width/length $<$ 0.5 and $\theta^{2}$ $<$ 0.015 deg$^{2}$. The background region was estimated using the ``ring background'' model \cite{Aharonian2005}. The results are presented in Table \ref{table_results}.\\

The significance calculations were done using equation 17 from Li and
Ma \cite{LiandMa}.  The 99$\%$ confidence level (C.L.) flux upper limits for
energy above 400 GeV are calculated using the Helene
method \cite{Helene}, assuming a power law energy spectrum of index -2.5.\\

\begin{table*}[th]
  \caption{Results on the X-ray binaries presented}
  \label{table_results}
  \centering
  \begin{tabular}{|c|c|c|c|c|c|c|c|c|}
  \hline
  Object      & Exposure & N$_{on}$ & N$_{off}$  & $\alpha$ & Significance  & 99$\%$ C.L. Upper limit (E$>$400 GeV) \\
               &   (h)    &          &            &          &   ($\sigma$)  &   (cm$^{-2}$s$^{-1}$)   [Crab Units]   \\   
  \hline      
   Low Mass X-ray binaries &          &            &          &  &  &                                                 \\
   \hline                           
    3A 1954+319   &      12.43    &     33   &   168      &   0.22   &     -0.6               &             9.66x$10^{-13}$ [1.0$\%$]          \\
    XTE J2012+381 &      13.91    &     90   &   305      &   0.25   &      1.5               &             2.45x$10^{-12}$ [2.5$\%$]          \\
    1A 0620-00    &       4.18    &     36   &   157      &   0.23   &      0.1               &             1.86x$10^{-12}$ [2.9$\%$]          \\
  \hline                                                                                                
   High Mass X-ray binaries   &          &            &          &  &  &                                      \\
  \hline                                                                                                          
    EXO 2030+375  &       6.69    &     32   &   115      &   0.27   &      0.2               &             2.22x$10^{-12}$ [2.2$\%$]           \\
    KS 1947+300   &       3.95    &     16   &    66      &   0.23   &      0.1               &             2.88x$10^{-12}$ [2.9$\%$]           \\
    SS 433        &      10.38    &      4   &   201      &   0.27   &     -0.8               &             1.50x$10^{-12}$ [1.5$\%$]          \\
    Cygnus X-1    &       9.75    &     32   &   142      &   0.25   &     -0.6               &             1.05x$10^{-12}$ [1.1$\%$]            \\
    Cygnus X-3    &      10.35    &     47   &   200      &   0.2    &      0.2               &             1.42x$10^{-12}$ [1.4$\%$]             \\
  \hline
  \end{tabular}
  \end{table*}

\section{Results and Discussion}

Results of our analysis on X-ray binaries are given in Table
\ref{table_results}, where N$_{on}$ is the number of events at the
source position, N$_{off}$ is the estimated number of background
events at the source position and $\alpha$ is the normalisation factor
between N$_{on}$ and N$_{off}$.  No gamma-ray emission has been
detected from any of these X-ray binaries observed by VERITAS. To our
knowledge, the analysis presents the first observations at TeV
energies of the binaries 3A 1954+319, XTE J2012+381, EXO 2030+375 and
KS 1947+300.\\

\noindent
\textit{3A 1954+319}

3A 1954+319 is one of the rare LMXB classified as a ``symbiotic X-ray
binary'' \cite{3A1954}\cite{Masseti}, composed of the slowest pulsar known accreting from the wind
of a M-type giant. This system has flaring behaviour in the
X-ray. Combining the 12.43 hours of observations on this object, we obtain a
99$\%$ C.L. flux upper limit (assuming steady flux) less than 1$\%$ of
the Crab Nebula. None of the VERITAS data was contemporaneous with
X-ray flares.\\

\noindent
\textit{XTE J2012+381}

XTE J2012+381 is a soft X-ray transient and is composed of a stellar
black hole candidate and a faint red companion star. The last major
X-ray outburst from this object was observed in May 1998 \cite{2012rxte,2012beppo}. We report a
99$\%$ C.L. flux upper limit for the quiescent phase of the transient
of $<$ 2.5$\%$ of the Crab Nebula flux. \\

\noindent
\textit{1A 0620-00}

1A 0620-00 is a soft X-ray transient and is composed of a black hole
with a K5 companion star which fills its Roche lobe \cite{a0620}. This object was
observed in outburst in 1974 \cite{a0620burst}. The 99$\%$ C.L. flux upper limit
calculation of $<$ 2.9$\%$ of the Crab Nebula flux is for the
quiescent phase of the transient. \\

\noindent
\textit{EXO 2030+375}

\begin{figure}[!t]
  \centering
  \includegraphics[angle=0,width=3in]{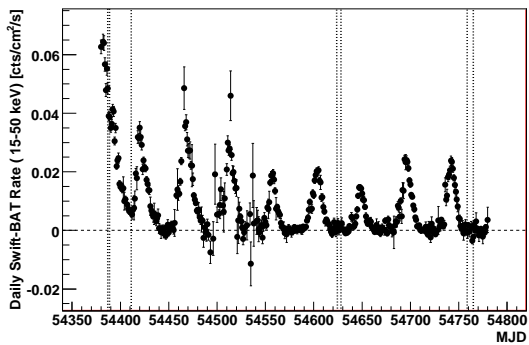}
  \caption{Swift daily light curve for EXO 2030+375 for the energy band 15-50 keV. The dotted lines show the time of the VERITAS observations.}
  \label{EXO_swift}
 \end{figure}

EXO 2030+375 is a X-ray pulsar accompanied by a B0 Ve star. This HMXB
has periodic X-ray outbursts at the periastron of the orbit and
showed a giant outburst in June 2006 \cite{Exo}\cite{Exo2}. Most of the VERITAS observations
fall in phases of the orbit where the X-ray emission is near its
minimum, see Figure \ref{EXO_swift}. The data taken on MJD 54389-54390
are only one run of 20 min. per night and in a single run we do not
have sufficient sensitivity for a meaningful limit. The analysis on
the 6.69 hours of data gives a 99$\%$ C.L. flux upper limit of $<$
2.2$\%$ of the Crab Nebula flux.\\

\noindent
\textit{KS 1947+300}

KS 1947+300 is a transient X-ray pulsar with a relatively high
magnetic field of 2.5x 10$^{13} G$ \cite{Kiziloglu2006} accompanied by
a B0 Ve star. The orbit of the system is very close to circular ($e$ =
0.033). This object is similar to A 0535+26, where TeV emission is
predicted from \cite{Orellana2007} during X-ray flares. The analysis
gives a 99$\%$ C.L. flux upper limit for a steady emission of $<$
2.9$\%$ of the Crab Nebula flux.\\


\noindent
\textit{SS 433}

\begin{figure}[!t]
  \centering
  \includegraphics[angle=90,width=3in]{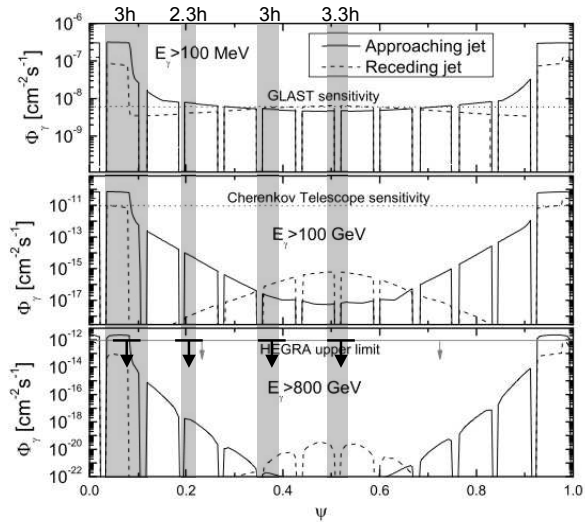}
  \caption{Gamma-ray fluxes at Earth as a function of the precessional phase. The lower panel shows the flux for energies above 800 GeV. The contributions of the two jets are shown separately: solid line for the approaching jet and dashed line for the receding one. Figure from \cite{Reynoso2008}. The shaded bands represent the regions in $\psi$ encompassed by the VERITAS observations. The total amount of observing time for each band is indicated at the top of the figure. VERITAS upper limits are shown by the black arrows.}
  \label{ss433}
 \end{figure}

SS 433 is a microquasar consisting of a black hole accreting from a
supergiant A star. In \cite{Reynoso2008}, a detailed hadronic model
for gamma-ray production in the microquasar is described. They make
predictions on the gamma-ray flux, at energies higher than 100 GeV and
800 GeV, arriving at Earth as a function of the different precession
cycle phases. They predict that the most favourable precessional
phases for gamma-ray detection are between 0.91 and 0.09, for a range
of $\sim$ 29 days, where the gamma-rays do not have to pass through
the equatorial disk along their path. Previous upper limits on the
gamma-ray flux have been presented by HEGRA from several hundreds
hours of observations, averaged on all precession phases, of
$<$ 8.93x10$^{-13}$cm$^{-2}$s$^{-1}$ at 99$\%$ C.L. for energies above
800 GeV ($<$ 3.2$\%$ of the Crab Nebula flux) \cite{RowellSS433}. Here,
we present a new upper limit on the gamma-ray flux calculated from
10.38 hours of observation with VERITAS of $<$ 1.5$\%$ of the Crab
Nebula flux at the source position. A subset of data was taken near a
most favourable time for potential gamma-ray detection. 2007 October
1st was the date where the precessional phase was 0, and 3 hours of
VERITAS observations were performed between October 5th to October
14th, corresponding to 4 to 13 days after the most favourable
time. Using only this data set, the total significance at the SS 433
position is -0.1$\sigma$ giving an upper limit of
$<$ 1.53x10$^{-12}$cm$^{-2}$s$^{-1}$ at 99$\%$ C.L. for energies above
800 GeV. This limit is shown in Figure \ref{ss433} with the limits for
the other precessional phases, where they compare with theoretical
predictions \cite{Reynoso2008} and HEGRA upper limit
\cite{RowellSS433}. \\

\noindent
\textit{Cygnus X-1}

\begin{figure}[!t]
  \centering
  \includegraphics[width=3in]{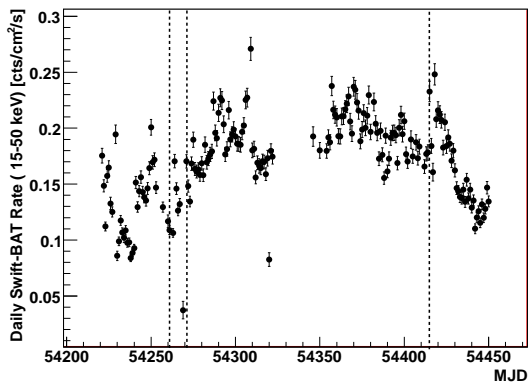}
  \caption{Swift daily light curve for Cygnus X-1 for the energy band 15-50 keV. The dotted lines show the period of VERITAS observations.}
  \label{cygx1_lc}
 \end{figure}

Cygnus X-1, one of the more intense X-ray sources in the sky, is composed
of a stellar black hole and a massive O9.7 Iab star \cite{cygx1ref}. It is also a
microquasar. Recently, MAGIC claimed a 3.2$\sigma$ detection after
trial correction of Cygnus X-1 during a X-ray flare
\cite{MAGIC_CygX1}. VERITAS analysis on 9.75 hours of data gives a
99$\%$ C.L. upper limit flux for a steady emission of $<$ 1.1$\%$ of
the Crab Nebula flux, which is consistent with MAGIC upper limit for steady emission
ranging between 1$\%$ and 2$\%$ \cite{MAGIC_CygX1}. None of VERITAS
data have been taken at the time of a X-ray flare, see Figure
\ref{cygx1_lc}. The last data set was taken the night after a high
X-ray count rate, although such intensity is not as high as one of the
major flares from this source. \\

\noindent
\textit{Cygnus X-3} 

Cygnus X-3 is also a microquasar with a Wolf-Rayet star companion \cite{cygx3}. We
present a 99$\%$ C.L. upper limit flux for a steady emission from
10.35 hours of observations of $<$ 1.4$\%$ of the Crab Nebula
flux. None of the observations coincide with X-ray flares.\\


VHE properties of X-ray binaries remain challenging. Future observations in the VHE range will bring more elements to the understanding of those objects. Multiwavelength campaigns, similar to the ones done for LS I+61 303 by VERITAS, SWIFT and RXTE \cite{andyls1}, or by VERITAS, FERMI and SUZAKU, will be really helpful to comprehend the full picture of the mechanisms in action in binaries. Some of the binaries presented here are part of Target of Opportunity program with VERITAS and future potential triggered observations will be useful to differentiate the behaviour during flares versus the quiescent state.\\

\section{Acknowledgements}

This research is supported by grants from the US Department of Energy, the US National Science Foundation, and the Smithsonian Institution, by FQRNT and NSERC in Canada, by Science Foundation Ireland, and by STFC in the UK. We acknowledge the excellent work of the technical support staff at the FLWO and the collaborating institutions in the construction and operation of the instrument.\\

\end{document}